\documentclass[12pt]{article}

\usepackage{amsmath}
\usepackage{amssymb}
\usepackage{slashed}
\usepackage[numbers,sort&compress]{natbib}
\usepackage{hyperref}
\usepackage{cleveref}
\crefname{equation}{Eq.}{Eqs.}
\crefname{figure}{Fig.}{Figs.}
\crefname{table}{Table}{Tables}
\crefname{section}{Section}{Sections}

\usepackage[letterpaper,margin=1in,bottom=1in]{geometry}
\usepackage{float} 
\usepackage{parskip} 
\usepackage{tabulary} 
\usepackage{color} 
\usepackage{soul} 
\usepackage{subfigure}
\usepackage{graphicx}
\usepackage[section]{placeins} 

\def\lhc2{LHC~Run~II}

\usepackage{cleveref}

\newcommand{\code}[1]{\texttt{#1}}

\def\.4{\vspace{-.5cm}}
\newcommand{\ifb}{~\textrm{fb}^{-1}}
\newcommand{\iab}{~\textrm{ab}^{-1}}

\def\beq{\begin{equation}}
\def\be{\begin{equation}}
\def\beqn{\begin{eqnarray}}
\def\ee{\end{equation}}
\def\eeq{\end{equation}}
\def\eeqn{\end{eqnarray}}

\author{
Amin Aboubrahim\footnote{Email: a.abouibrahim@northeastern.edu}~\ and 
Pran Nath\footnote{Email: p.nath@northeastern.edu}\\~\\
Department of Physics, Northeastern University,
Boston, MA 02115-5000, USA
}

\title{Naturalness, the Hyperbolic Branch and Prospects for the Observation of  Charged Higgs at  High Luminosity LHC and 27 TeV LHC}

\begin{document}
\maketitle
\date

\textbf{Abstract: } 
One of the early criterion proposed for naturalness was a relatively small MSSM Higgs mixing parameter $\mu$ 
with $\mu/M_Z$ of the order of a few.
 A relatively small $\mu$ may lead to heavier Higgs masses ($H^0, A, H^{\pm}$ in MSSM) which 
  are significantly
 lighter than other scalars such as squarks. 
Such a situation is realized on the hyperbolic branch of radiative breaking of the electroweak symmetry.
In this analysis we construct supergravity unified models with relatively small $\mu$  in the sense described above and
discuss the search for the charged Higgs boson $H^{\pm}$ at HL-LHC and HE-LHC where we also carry out a relative comparison of the discovery potential of the two
using the decay channel $H^{\pm} \to \tau \nu$.
It is shown  that an analysis based on the traditional linear cuts on signals and backgrounds is not very successful in extracting the signal while, 
in contrast, machine learning techniques such as boosted decision trees prove to be far more effective. 
Thus it is shown that  models  not discoverable with the conventional cut analyses become discoverable with machine 
learning techniques. Using boosted decision trees we  consider several  benchmarks and analyze the potential 
for their $5\sigma$ discovery at the 14 TeV HL-LHC and at  27 TeV HE-LHC. It is shown that while the ten benchmarks considered
with the charged Higgs boson mass in the range 373 GeV-812 GeV are all discoverable at HE-LHC, only four of the ten 
with 
Higgs boson masses in the range 373 GeV-470 GeV are discoverable at HL-LHC. Further, while the model points discoverable at both
HE-LHC and HL-LHC would require up to 7 years of running time at HL-LHC, they could all  be discovered in a period of few months  
at HE-LHC. The analysis shows  that a transition  from HL-LHC to HE-LHC when technologically feasible  would expedite 
the discovery of the charged Higgs for the benchmarks considered in this work. We note that the observation of a charged Higgs 
boson with mass in the range indicated would lend support to the idea of naturalness defined by a relatively small $\mu$ and  further, it will 
lend support to radiative breaking of the electroweak symmetry occurring on the hyperbolic branch.

\newpage

\section{Introduction \label{sec:intro}}

The discovery of the Higgs boson~\cite{Englert:1964et, Higgs:1964pj, Guralnik:1964eu} in 2012 by the ATLAS and CMS 
collaborations~\cite{Chatrchyan:2012ufa, Aad:2012tfa}
was a landmark and contains clues to the 
nature of physics beyond the standard model. Thus  in the standard model the Higgs boson can be as large as 800 GeV
while in supergravity (SUGRA) grand unified models~\cite{msugra} (for a review see~\cite{Nath:2016qzm})
 it is predicted to lie below 130 GeV~\cite{Akula:2011aa}.
Further, the Higgs boson is discovered with
a mass of $\sim 125$ GeV  exhibiting the fact that the supergravity limit of 130 GeV is respected.  However, within supersymmetry (SUSY) the tree level Higgs boson mass is predicted to lie below the Z-boson mass, which indicates that the loop corrections are
rather large which in turn points to the {size} of weak scale supersymmetry lying 
in the several TeV region~\cite{Akula:2011aa}.
The {large size} of weak scale supersymmetry makes the observation of sparticles more difficult. Further, with larger sfermion
masses efficient annihilation of dark matter particles becomes more difficult and typically requires 
coannihilation~\cite{Griest:1990kh} to be consistent with the WMAP~\cite{Larson:2010gs} and Planck data~\cite{Aghanim:2018eyx}.
Coannihilation in turn implies that the decay of the next-to-lightest supersymmetric particle (NLSP) will produce soft final states in models with $R$-parity which makes the  detection of supersymmetry also more difficult. 
These constraints are softened in models where
 the wino or the higgsino content of the neutralino is significant as shown for some of the models discussed in section 3.

 It should be noted that the {large size} of weak scale supersymmetry resolves some of the
problems associated with low scale supersymmetry. One of these concerns taming the CP phases that arise in the soft breaking 
sector of supersymmetry and can produce large electric dipole moments in conflict with the experimental limits that currently exist.
One of the ways to control them is the cancellation mechanism~\cite{cancellation,Ibrahim:2007fb}.  
However, if the sfermion masses lie in the several TeV region,
the CP phases would be automatically controlled~\cite{Nath:1991dn,Kizukuri:1992nj}. 
 In unified models based on supersymmetry one persistent problem relates
to the dangerous proton decay arising from baryon and lepton number violating  dimension five operators.  However, such 
operators are signficantly suppressed if the scalar masses lie in the several TeV region~\cite{Liu:2013ula,Nath:2006ut}. 
Another problem that finds resolution
if the weak scale is large relates to the so-called gravitino problem, in that a gravitino with a mass larger than 10 TeV will
decay  early enough not to interfere with big bang nucleosynthesis~\cite{Aboubrahim:2017wjl}.
It is also quite remarkable that supergravity models with sizable  scalar masses in the range of several
TeV  are consistent with the unification of gauge couplings~\cite{Aboubrahim:2017wjl}.
Thus the case for supersymmetry is stronger as a consequence of the discovery of the standard model-like Higgs boson~\cite{Nath:2018rqn,Nath:2015dza}.

One of the signatures of supersymmetric models is the existence of at least two Higgs doublets which leads to two more 
  neutral Higgs bosons, one CP even $H^0$ and one CP odd $A^0$, and two charged Higgs $H^{\pm}$. Thus an  indication of the  
existence of new physics beyond the standard model and an indirect support for supersymmetry can also come via
  discovery of one or more heavier Higgs bosons beyond the Higgs boson of the standard model. 
In supergravity unified models radiative breaking of the electroweak symmetry leads in general to two branches, one is the so-called
  ellipsoidal branch and the other is the hyperbolic branch~\cite{Chan:1997bi,Chattopadhyay:2003xi,Akula:2011jx}  
  (for related works see~\cite{Feng:1999mn,Baer:2003wx,Feldman:2011ud,Ross:2017kjc}).   
    On the hyperbolic branch the MSSM Higgs mixing parameter $\mu$ can be relatively small
  with $\mu/M_Z$ order a few while the squarks masses can lie in the several TeV region and provide the desired loop correction 
  to the standard model-like Higgs boson to lift its tree level value from below $M_Z$ to its experimentally measured value.  
  However, a relatively small $\mu$ points to  relatively light heavier Higgs bosons (relative to the the squark masses)
  and thus candidates for discovery at colliders. 
   In this 
  work we focus on the potential of the LHC to discover the charged Higgs within SUGRA models which 
  are high scale models where SUSY is broken by gravity mediation.   Currently LHC is in its second phase 
  which we might call {LHC Run 2} after the very successful {LHC Run 1} which discovered the Higgs boson. The {LHC Run 2}  will run till the end of 2018 and each detector will collect about 150 fb$^{-1}$ of data. It will then shut down for  
  two years for an   upgrade to {LHC Run 3} which will resume its run in  the period 2021-2023 and it is expected to collect
  300 fb$^{-1}$ of additional data. After that there will be a major upgrade to {LHC Run 4} in the period 2023-2026 
   with an upgrade to $\sqrt s= 14$ TeV and to high luminosity. It is expected that {LHC Run 4} will resume its operations
   in 2026 and run for ten years at the end of which an integrated luminosity of 3000 fb$^{-1}$ will be achieved.
   
Assuming that the discovery of a sparticle or a heavier Higgs is made at the LHC, a full exploration of the spectrum of 
   the sparticle masses and of  the Higgses will require a higher energy machine and there are dedicated groups investigating 
   this possibility.  One of the possibilities discussed is a 100 TeV proton-proton collider at CERN. This would require 
   a  100 km circular ring in the lake Geneva basin. Another possibility discussed is that of a 100 TeV proton-proton collider in 
   China~\cite{Arkani-Hamed:2015vfh,Mangano:2017tke}.
    However, a new possibility has recently been discussed 
    which is a 27 TeV proton-proton collider~\cite{Benedikt:2018ofy,Zimmermann:2018koi,HE-LHC-1,HE-LHC-2,cern-report}     
    which can be built
   in the existing CERN ring with 16 Tesla superconducting magnets using the FCC technology. Such a machine 
   will operate with a luminosity of $2.5\times 10^{35}$ cm$^{-2}$s$^{-1}$  and collect up to 15  ab$^{-1}$ of data. 
       In a recent work~\cite{Aboubrahim:2018bil}     
     an analysis on the potential for the discovery of supersymmetry at  HL-LHC vs
      HE-LHC was carried out.  In this work we carry out a similar analysis to discuss the potential for the discovery of charged Higgs at the HL-LHC vs
      HE-LHC.  Here we make a further comparison of  the conventional linear cuts vs  machine learning tools for the
      discovery. Specifically we use in our analysis boosted decision trees (BDT) and show that  some of the models which are undiscoverable at
      HL-LHC using the conventional linear cut analysis can be discovered using boosted decision tree technique. We carry out a
      similar analysis for HE-LHC. Here we show that HE-LHC is much more powerful for the discovery of the charged Higgs
      than HL-LHC.   
      
The outline of the rest of the paper is as follows: In section~\ref{sec2} we give an overview of the Higgs sector in the MSSM, in section~\ref{sec3} we give a review of the {hyperbolic branch} of radiative breaking of electroweak symmetry and in section~\ref{sec4} we describe the SUGRA model and the benchmark points used in this analysis satisfying the Higgs boson mass and the dark matter relic density. In section~\ref{sec5} we describe the production modes of the charged Higgs in association with a top [bottom] quark in the four and five flavour schemes and give the respective production cross-sections and charged Higgs branching ratios for the benchmark points. The codes used for simulation of signal and background samples are described in section~\ref{sec6} along with the selection criteria used to study the discovery potential of the charged Higgs in its $\tau\nu$ decay. Also the two methods for signal analysis, 
 linear cut-based and boosted decision trees, are explained and compared. In section~\ref{sec7} we discuss dark matter direct detection for the SUGRA benchmark points and in section~\ref{sec8} we give conclusions.

\section{The Higgs sector in the MSSM \label{sec2}}

In the minimal supersymmetric standard model (MSSM), the Higgs sector contains two Higgs doublets $H_d$ and $H_u$,
\begin{align}
   H_d &= \begin{pmatrix}
           H^0_d \\
           H^{-}_d \\
         \end{pmatrix}
         ~~\text{and} ~~
    H_u = \begin{pmatrix}
           H^+_u \\
           H^{0}_u \\
         \end{pmatrix},
  \end{align}
with opposite hypercharge which ensures the cancellation of chiral anomalies. Here $H_d$ gives mass to the down-type quarks and 
the leptons while $H_u$ gives mass to up-type quarks. 
The Higgs potential in the MSSM arises from three sources: the $F$ term of the superpotential, the $D$ terms containing the quartic Higgs interaction and the soft SUSY breaking Higgs mass squared, $m_{H_d}^2$ and  $m_{H_u}^2$, and the bilinear $B$ term. The full CP-conserving Higgs scalar potential can be written as~\cite{Djouadi:2005gj}
\begin{align}
V_H&=(|\mu|^2+m^2_{H_d})|H_d|^2+(|\mu|^2+m^2_{H_u})|H_u|^2-\mu B\epsilon_{ij}(H^i_u H^j_d + \text{h.c.}) \nonumber \\ 
&+ \frac{g_1^2+g_2^2}{8}(|H_d|^2-|H_u|^2)^2+\frac{1}{2}g_2^2|H_d^{\dagger}H_u|^2 + \Delta V_{\text loop}\,,
\end{align}
where $\mu$ is the Higgs mixing parameter appearing in the superpotential term $\mu \hat{H}_u\cdot \hat{H}_d$. 
 Minimization of the potential which preserves color and charge gives two constraints one of which can be used to determine 
  $\mu$  up to a sign, and the other to eliminate $B$ in favor of $\tan\beta =v_d/v_u$. 
The neutral components of the Higgs doublets can be expanded  around their VEVs so that
\begin{align}
H^0_d&=\frac{1}{\sqrt{2}}(v_d+\phi_d + i \psi_d),\nonumber\\
H^0_u&=\frac{1}{\sqrt{2}}(v_u+ \phi_u  + i \psi_u)\,.
\end{align}

After spontaneous breaking the mass diagonal charged and CP odd neutral Higgs fields are given by
\begin{align}
H^{\pm}&=-H_d^{\pm}\sin\beta+H_u^{\pm}\cos\beta, \\
A&=-\psi_d\sin\beta+ \psi_u\cos\beta\,. 
\end{align} 
The charged  and the CP odd neutral Higgs boson masses at the tree level
are given by 
\begin{equation}
m^2_{H^{\pm}}=m^2_A+m^2_W~~; ~~m_A^2= \frac{2B\mu}{\sin(2\beta)}\,.
\label{chargedhiggsmass}
\end{equation}
In the MSSM, the couplings of the charged Higgs boson to up-type fermions go as $\cot\beta$ whereas the coupling goes as $\tan\beta$ for down-type fermions. In this paper we will be looking at the leptonic decays of the charged Higgs boson and so enhancing this channel requires larger $\tan\beta$ values. This has become increasingly difficult for low masses of the charged Higgs since exclusion limits tend to be more severe for the high $\tan\beta$-low mass regime as will explain in the coming sections. 
The Higgs sector of the MSSM is similar to the 2HDM-type II with some differences such as the SUSY QCD corrections which are present only for the MSSM case.  For reviews on the Higgs sector of the MSSM and the 2HDM see Refs.~\cite{Gunion:1989we,Gunion:1992hs,Branco:2011iw}.

\section{Naturalness and the hyperbolic branch of radiative breaking\label{sec3}}
Issues of naturalness arise in the context of radiative breaking of the electroweak symmetry where one of the stability conditions is given by 
\begin{align}
\mu^2&=\frac{m^2_{H_u}\sin^2\beta-m^2_{H_d}\cos^2\beta}{\cos2\beta}-\frac{M^2_Z}{2} + \delta \mu^2 \,, 
\label{minconditions}
\end{align}
where $\delta \mu^2$ is the loop correction~\cite{Arnowitt:1992qp}. To illustrate the origin of the hyperbolic branch we consider the 
case of universal boundary conditions given by $m_0, A_0, m_{1/2}, \tan\beta$ and sign$(\mu)$ where $m_0$ is the universal
scalar mass, $A_0$ is the universal trilinear coupling, $m_{1/2}$ is the universal gaugino mass all at the GUT scale and $\tan\beta$
as defined earlier (the analysis of the hyperbolic branch 
for the non-universal case can be found in~\cite{Liu:2013ula}). 
Thus for the universal boundary conditions at the GUT scale we  may write this equation in terms of the 
parameters at the GUT scale~\cite{Nath:1997qm,Chan:1997bi} so that
 \beqn
 \mu^2  &=& -\frac{1}{2}M_Z^2 +  m^2_0  C_1+ A^2_0 C_2
+ m^2_{1/2} C_3+ m_{1/2}
A_0 C_4+ \Delta \mu^2_{\rm loop}~,
\label{1.6}
\eeqn
where 
\beqn \label{1.61}
C_1&=&\frac{1}{\tan^2\beta-1}\left(1-\frac{3 D_0-1}{2}\tan^2\beta\right)~,\\
\label{1.61a}
 C_2&=&\frac{\tan^2\beta}{\tan^2\beta-1}k~,\\
 \label{1.61b}
C_3&=&\frac{1}{\tan^2\beta-1}\left(g- e\tan^2\beta \right)~, \\
C_4&=&-\frac{\tan^2\beta}{\tan^2\beta-1}f~,
\label{1.8}
\eeqn
 and 
 $D_0(t)$ is  defined by 
\beqn
 D_0(t)= \left(1+ 6 Y_0 F(t)\right)^{-1}~.
 \label{d0EQ}
 \eeqn
Here $Y_0 =h^2_t(0)/(4\pi^2)$, where $h_t(0)$ is the top Yukawa coupling at the GUT scale $M_G$,  
$F(t) = \int_0^t E(t^\prime) dt^\prime$, where  
$E(t)=\left(1 + \beta_3 t\right)^{16/3b_3} \left(1+ \beta_2 t\right)^{3/b_2} \left(1+ \beta_1t\right)^{13/9 b_1}$.
Here 
$\beta_i = \alpha_i(0) b_i/(4\pi)$ and $b_i=(-3, 1, 11)$ for $SU(3), SU(2)$ and $U(1)$
 and $t= \ln \left(M_G^2/Q^2\right)$ where $Q$ is the
 renormalization group point.  Our normalizations are such that $\alpha_3(0) = \alpha_2(0) = \frac{5}{3} \alpha_1(0)
 =\alpha_G(0)$. The functions  $e,f,g,k$ are as defined in~\cite{Ibanez:1984vq}.  
An interesting aspect of Eq.~(\ref{1.6}) is that it relates $\mu$,  which enters in the superpotential, to the soft breaking terms. This raises the issue of what the size of $\mu$ is. 
One very obvious choice is the following: the radiative breaking equation is supposed to generate masses for the vector bosons
$W$ and $Z$. Thus a reasonable choice is to have $\mu/M_Z$ which is order few, i.e., $\mu/M_Z\sim (1-5)$ which was essentially
the criterion of naturalness adopted in~\cite{Chan:1997bi}.  We note here that small $\mu$ models have been investigated 
quite extensively recently (see, e.g.,~\cite{Baer:2018avn,Baer:2018hpb,Baer:2018rhs} and the references therein).

  It is important to note that a relatively small $\mu$ discussed above does not necessarily imply that $m_0$ need be small.
  To illustrate this point, it is useful to exhibit the  underlying geometry of the radiative breaking equation. Thus, as is well known,
   both the tree value of $\mu^2$  given by Eq.~\eqref{1.6} and the loop correction $\Delta \mu^2_{\rm loop}$,
   have significant dependence on the  renormalization group scale. 
   Their sum, however, is relatively insensitive to the changes in the renormalization group scale~\cite{Chan:1997bi}.  
 Thus suppose we go to the renormalization group point where the loop correction is small, and here we may simply consider the
 tree formula for $\mu^2$. 
  It was seen in~\cite{Chan:1997bi} that while for part of the parameter space of supergravity models 
    all the $C_i, ~(i=1-4)$ are positive,
  there are regions of the parameter space where $C_1$ can vanish or even turn negative. Reference to 
  Eq.~(\ref{1.6})  shows that for the case when $C_1=0$, $\mu^2$ becomes independent of $m_0$ (this is the so-called focal point region).
  Similarly when $C_1<0$, one finds curves in the $m_0-A_0$ plane where the sum of the contributions to $\mu^2$
    involving $m_0$ and $A_0$ vanish.  This is what one may call the focal curve region~\cite{Akula:2011jx}. 
    The same idea extends to focal 
    surfaces.
     In these regions $\mu$ and one or more of the  soft parameters are uncorrelated. Thus, for example, $m_0$ and $A_0$ can be chosen without affecting $\mu$ in the focal curve region.
 
  We wish to note here that concepts such as  naturalness are generally invoked only in the context of an incomplete theory which means
  that some parameters in the theory are unknown and one must  make reasonable choices for them for investigating the theory. 
  However, choices which might appear unreasonable need not be so if they are dictated by the internal constraints of 
  the more complete theory, which means that no doors need be closed when working with an incomplete theory. 
  Thus in the analysis below we will consider two ranges for $\mu$:  one which fits the criteria 
   discussed above, i.e., $\mu/M_Z$ in the range (1-5) and for the other we will  step   outside this range.
     We note that the analysis above shows that the observation of one of the heavier Higgs bosons $H^0, A^0, H^{\pm}$
    with masses much less than $m_0$ would point to radiative breaking of the electroweak symmetry on the 
    hyperbolic branch and further if these Higgs bosons are observed with masses in the few hundred GeV range, that would lend
     support to naturalness defined by small $\mu$.

\section{ SUGRA model benchmarks \label{sec4}}
The focus of this work is to explore the potential of HL-LHC and HE-LHC for discovering a charged Higgs boson in 
 a class of high scale models, specifically SUGRA models consistent with the experimental constraints on 
 the light Higgs mass at $\sim 125$ GeV. The analysis is done  under the constraints of $R$-parity so the LSP is stable. 
 Further, in a large part of the  parameter space in SUGRA models it is found that the LSP is also the lightest neutralino 
 and thus a candidate for dark matter and the models are thus subject to the constraints that they be consistent with the observed 
 amount of cold dark matter so that~\cite{Aghanim:2018eyx},
\begin{equation}
\Omega h^2=0.1198\pm 0.0012.
\label{relic}
\end{equation}
 Consistency with Eq.~(\ref{relic}) would require non-universalities  in the gaugino sector. Further, from 
   Eq.~(\ref{chargedhiggsmass})  we note that  the charged Higgs mass depends on the mass of the CP odd neutral Higgs $A^0$ 
 which in turn depends on the Higgs mixing parameter, $\mu$ and $\tan\beta$. We wish to have  charged Higgs masses 
 in the range $\sim (300- 800)$ GeV, which requires 
 non-universalities in  $m_{H_d}$ and  $m_{H_u}$ at the grand unification scale. Including the non-universalities in the gaugino sector 
 (for recent works see \cite{nonuni2}) and in the Higgs masses (for recent works see \cite{NU}  and for a review see \cite{Nath:2010zj}),
 the extended SUGRA parameter space  at the GUT scale  is given  by
\begin{equation}
m_0, ~~A_0, ~~m_1, ~~m_2, ~~m_3, ~~m^0_{H_u}, ~~m^0_{H_d}, ~~\tan\beta, ~~\text{sgn}(\mu),
\label{sugra}
\end{equation}    
where $m_1,  m_2, m_3$ are the masses of the $U(1)$, $SU(2)$, and $SU(3)_C$ gauginos, and $m^0_{H_u}$ and $m^0_{H_d}$ are the 
masses of the up and down Higgs bosons all at the GUT scale.
In Table~\ref{tab1} we exhibit ten benchmarks which lead to Higgs masses and sparticle masses in the ranges not excluded.
We note here that satisfaction of the relic constraint requires  coannihilation in the models we consider. Coannihilation has been 
considered in a variety of recent works~\cite{Kaufman:2015nda,Nath:2016kfp,Aboubrahim:2017aen}
where with stop, gluino and stau coannihilations were considered. Here as in the analysis of~\cite{Aboubrahim:2017wjl} 
the coannihilating particle is the lightest chargino, $\chi^{\pm}_1$ which implies that the chargino and the LSP mass gap must be small,
i.e., 
$$(m_{\tilde{\chi}^{\pm}_1}-m_{\tilde{\chi}^0_1})\ll m_{\tilde{\chi}^{0}_1}.$$
The mass spectrum of the model is calculated using \code{softSUSY}~\cite{Allanach:2001kg,Allanach:2016rxd} 
and the relic density is evaluated using \code{micrOMEGAs}~\cite{Belanger:2014vza}. Taking the computational uncertainties in the codes into consideration, the light Higgs mass constraint is taken to be $125\pm 2$ GeV and the relic density as $\Omega h^2 < 0.126$. 
For the case when the cold dark matter constitutes only a fraction of dark matter, one would have multi-component 
dark matter~\cite{Feldman:2010wy} (one such recent possibility is the ultralight dark axion, see, e.g.,~\cite{Hui:2016ltb,Halverson:2017deq}). 

\begin{table}[H]
\begin{center}
\begin{tabulary}{0.85\textwidth}{l|CCCCCCCC}
\hline\hline\rule{0pt}{3ex}
Model & $m_0$ & $m^0_{H_d}$ & $m^0_{H_u}$ & $A_0$ & $m_1$ & $m_2$ & $m_3$ & $\tan\beta$ \\
\hline\rule{0pt}{3ex}  
\!\!(a)& 5032 & 434 & 9034 & -12429 & 960 & 285 & 3395 & 8 \\
(b)  & 4232 & 1077 & 8480 & -11859 & 709 & 250 & 3358 & 10 \\
(c)  & 4280 & 921 & 7702 & -10400 & 485 & 269 & 2968 & 10 \\
(d)  & 5311 & 1512 & 9854 & -14126 & 484 & 272 & 3709 & 11 \\
(e)  & 4636 & 1105 & 8093 & -11034 & 400 & 252 & 3090 & 11 \\
(f)  & 5785 & 1402 & 8577 & -12322 & 395 & 238 & 2477 & 12 \\
(g)  & 3820 & 1895 & 7395 & -10390 & 585 & 318 & 2843 & 15 \\
(h)  & 5800 & 2159 & 8365 & -12200 & 686 & 425 & 2280 & 16 \\
(i)  & 7150 & 2001 & 9582 & -14300 & 380 & 233 & 2250 & 15 \\
(j)  & 3399 & 2335 & 7082 & -9823 & 472 & 259 & 2900 & 18 \\
\hline
\end{tabulary}\end{center}
\caption{Input parameters for the SUGRA benchmark points used in this analysis. All masses are in GeV.}
\label{tab1}
\end{table}

\begin{table}[H]
\begin{center}
\begin{tabulary}{1.3\textwidth}{l|CCCCCCCCC}
\hline\hline\rule{0pt}{3ex}
Model  & $h^0$ & $\mu$ & $\tilde\chi_1^0$ & $\tilde\chi_1^\pm$ & $\tilde t$ & $\tilde g$ & $A^0\sim H^0$ & $H^{\pm}$ & $\Omega^{\rm th}_{\tilde{\chi}^0_1} h^2$ \\
\hline\rule{0pt}{3ex} 
\!\!(a) & 123.0 & 232 & 153.5 & 156.6 & 3177 & 7140 & 364 & 373 & $1.06\times 10^{-3}$  \\
(b) & 123.1 & 257 & 139.0 & 140.9 & 2832 & 7027 & 408 & 416 & $6.76\times 10^{-4}$  \\
(c) & 123.2 & 429 & 175.8 & 177.9 & 2982 & 6282 & 432 & 439 & $9.32\times 10^{-4}$  \\
(d) & 123.0 & 422 & 170.7 & 172.5 & 3079 & 7747 & 463 & 470 & $8.71\times 10^{-4}$  \\
(e) & 123.2 & 688 & 157.3 & 168.7 & 3185 & 6536 & 504 & 511 & $1.34\times 10^{-3}$  \\
(f) & 123.2 & 863 & 161.9 & 173.6 & 2640 & 5419 & 561 & 567 & $1.29\times 10^{-3}$   \\
(g) & 123.3 & 373 & 213.2 & 216.9 & 2408 & 6006 & 611 & 616 & $1.55\times 10^{-3}$   \\
(h) & 123.1 & 1039 & 295.5 & 343.4 & 2451 & 5018 & 652 & 657 & $1.24\times 10^{-1}$   \\
(i) & 123.3 & 1587 & 160.7 & 181.7 & 2954 & 5030 & 696 & 701 & $9.40\times 10^{-2}$  \\
(j) & 123.8 & 339 & 163.7 & 166.5 & 2552 & 6096 & 808 & 812 & $9.18\times 10^{-4}$   \\
\hline
\end{tabulary}\end{center}
\caption{The Higgs boson ($h^0$) mass, the $\mu$ parameter, the heavy Higgses ($A^0, H^0, H^{\pm}$) and some relevant sparticle masses, and the relic density for the benchmarks  of Table~\ref{tab1}. All masses are in GeV. }
\label{tab2}
\end{table}
Table~\ref{tab2} exhibits the light and heavy Higgs masses and  the masses of the electroweakinos, the stop and gluino masses along 
with the $\mu$  value and the relic density. 
Here $\mu$ is in the range $(200-1600)$ GeV. 
For small $\mu$, the neutralino has a larger Higgsino content leading to an efficient annihilation of these neutralinos in the
early universe. Thus the relic density here can be significantly smaller than indicated by Eq.~(\ref{relic}).

\section{Charged Higgs production in association with a top (and bottom) quark \label{sec5}}
The production of the charged Higgs boson has been extensively studied theoretically and experimentally for most mass ranges. Thus we consider a charged Higgs as light when its mass is much smaller than that of the top quark. Such a particle has been excluded by Tevatron~\cite{Aaltonen:2009ke} and LEP~\cite{Abbiendi:2013hk} for the entire $\tan\beta$ range. For moderate mass ranges, $m_{H^{\pm}}\sim 150-170$ GeV, no firm experimental analysis exists because of the  absence of theoretical studies of the signal  that include  important width effects
 and a full amplitude analysis 
 for $pp\rightarrow H^{\pm}W^{\mp}b\bar{b}$ is needed~\cite{deFlorian:2016spz}. Thus in the exclusion limits for the charged Higgs given by ATLAS and CMS one finds a mass gap as noted above.
  In this analysis we consider heavy charged Higgs, i.e., $m_{H^{\pm}} > m_t$.
In this region 
 ATLAS and CMS~\cite{Aaboud:2018cwk,Aaboud:2018gjj,Aaboud:2017qph,ATLAS:2016qiq,ATLAS:2016grc,Vischia:2016psk,CMS:2018ect} 
have
excluded masses up to 1100 GeV for $\tan\beta\sim 60$ while masses up to 400 GeV are excluded for $\tan\beta<2$ for the channel $H^{\pm}\rightarrow \tau\nu$ except for the small gap mentioned previously. In the same channel, all masses up to 600 GeV are excluded for $\tan\beta\sim 50$. The more stringent constraints on the charged Higgs mass come from constraints on the CP odd Higgs~\cite{CMS:2017epy,CMS:2016rjp,Aaboud:2017sjh,Aaboud:2016cre}. In the $A\rightarrow\tau\tau$ channel, all masses up to 1000 GeV are excluded for $\tan\beta<6$, while masses up to 1500 GeV are excluded for $\tan\beta\gtrsim 45$. Low masses $(300<m_A<500)$ are only allowed for $\tan\beta<10$ whereas higher $\tan\beta$ values require larger masses. Constraints on the CP odd Higgs translates into constraints on the charged Higgs mass according to Eq.~(\ref{chargedhiggsmass}). {We note that while the Higgs mass relations given by 
    Eq. (\ref{chargedhiggsmass}) are tree level MSSM mass relations, in the actual analysis the Higgs masses
    calculated at full one-loop order with \code{SoftSUSY} are used.}    
The masses of the CP odd and charged Higgses of the ten benchmark points of Table~\ref{tab2} are still allowed along with the masses of the 
charginos and neutralinos which belong to a compressed spectrum~\cite{Dutta:2015exw,Berggren:2015qua,Berggren:2016qjh,LeCompte:2011fh} (for experimental searches on compressed spectra, see Refs.~\cite{Khachatryan:2016mbu,Khachatryan:2016pxa,MORVAJ:2014opa,Aaboud:2017leg}). \\
The largest production mode of the charged Higgs at hadron colliders is the one that proceeds in association with a top quark (and a low transverse momentum $b$-quark),
\begin{eqnarray}
pp &\longrightarrow& t[b]H^{\pm}+X.
\label{pp}
\end{eqnarray}
This production mode can be realized in two schemes, namely, the four and five flavour schemes (4FS and 5FS, respectively), where in the former, the $b$-quark is produced in the final state and  in the latter it is considered as part of the proton's sea of quarks and folded into the parton distribution functions (PDF). This difference between 4FS and 5FS comes about mainly due to the collinear splitting of an  incoming gluon into a $b\bar{b}$ pair resulting in large logarithms which can be absorbed into the DGLAP equations~\cite{Altarelli:1977zs} thus making up the 5FS approach. Here, the final state $b$-quark is assumed massless and has low transverse momentum. Also, the virtual $b$-quark has a zero virtuality (i.e., $m\approx 0$). The cross-sections of the two production modes
\begin{eqnarray}
q\bar{q}, gg &\longrightarrow& tbH^{\pm} ~~~~ \text{(4FS),} \nonumber \\
gb &\longrightarrow& tH^{\pm} ~~~~~ \text{(5FS),}
\label{parton}
\end{eqnarray}
are evaluated at next-to-leading order (NLO) in QCD with \code{MadGraph\_aMC@NLO-2.6.3}~\cite{Alwall:2014hca} using \code{FeynRules}~\cite{Alloul:2013bka} UFO files~\cite{Degrande:2011ua,Degrande:2014vpa} for the Type-II two Higgs doublet model (2HDM). The simulation is done at fixed order, i.e., no matching with parton shower. For NLO accuracy at both fixed order and with parton shower matching see Ref~\cite{Degrande:2015vpa}. The couplings of the 2HDM are the same as in the MSSM, but when calculating production cross-sections in the MSSM, one should take into account the SUSY-QCD effects. In our case, as one can see from Table~\ref{tab2}, gluinos and stops are rather heavy and thus their loop contributions to the cross-section is very minimal. In this case, the 2HDM is the decoupling limit of the MSSM and this justifies using the 2HDM code to calculate cross-sections. For the 5FS, the bottom Yukawa coupling is assumed to be non-zero and normalized to the on-shell running $b$-quark mass which is also calculated with \code{MadGraph} at the hard scale of the process. Eq.~(\ref{parton}) shows the parton level subprocesses responsible for the production of a charged Higgs in association with a top quark at leading order (LO). Note that in Eqs.~(\ref{pp}) and~(\ref{parton}), $t$ may refer to a top or antitop, and $b$ to bottom or antibottom depending on the sign of the charged Higgs. In the 5FS, the process is initiated via gluon-$b$-quark fusion while in the 4FS it proceeds through either quark-antiquark annihilation (small contribution) or gluon-gluon fusion. In fact, the NLO cross-section of the 5FS process contains an $\mathcal{O}(\alpha_s)$ correction which includes, at tree-level, the 4FS processes. At finite order in perturbation theory, the cross-sections of the two schemes do not match due to the way the pertubative expansion is handled but one expects to get the same results for 4FS and 5FS when taking into account all orders in the perturbation. In order to combine both estimates of the cross-section, we use the Santander matching criterion~\cite{Harlander:2011aa} whereby 
\begin{equation}
\sigma^{\rm matched}=\frac{\sigma^{4\rm FS}+\alpha\sigma^{5\rm FS}}{1+\alpha},
\label{matched}
\end{equation}  
with $\alpha=\ln\left(\frac{m_{H^{\pm}}}{m_b}\right)-2$. The matched cross-section of the inclusive process lies between the 4FS and 5FS values but closer to the 5FS value owing to the weight $\alpha$ which depends on the charged Higgs mass. The uncertainties are combined as such, 
\begin{equation}
\delta\sigma^{\rm matched}=\frac{\delta\sigma^{4\rm FS}+\alpha\delta\sigma^{5\rm FS}}{1+\alpha}.
\end{equation} 
Table~\ref{tab3} shows the NLO cross-sections for the charged Higgs top-associated production in the 4FS and 5FS for two center-of-mass energies, 14 TeV and 27 TeV. The matched cross-sections along with the uncertainties are given. A factor of $\sim 5$ to 8 increase in the production cross-section is seen when going from 14 TeV to 27 TeV. In the 5FS, the cross-sections of the ten benchmark points are up to $\sim 2$ times larger than those obtained with the 4FS due to the presence of an additional coupling factor in the latter. The renormalization and factorization scales are chosen as the hard scale of the process with $\mu_R=\mu_F=\frac{1}{3}(m_t+\bar{m}_b+m_{H^{\pm}})$, with $\bar{m}_b$ being the running $b$-quark mass. The charged Higgs production cross-section is proportional to
\begin{equation}
\sigma_{H^{\pm}}\varpropto m^2_t\cot^2\beta+m^2_b\tan^2\beta\pm 2m_t m_b,
\end{equation} 
which points to a dip in the cross-section around $\tan\beta=8$ to 9. The cross-section increases for $\tan\beta\lesssim 8$ and $\tan\beta > 10$. Table~\ref{tab3} demonstrates the fall of the cross-section with the charged Higgs mass but it also shows the effect of $\tan\beta$. For example, in going from point (f) to point (g), one can observe a rise in the cross-section due to the increase in $\tan\beta$. 

\begin{table}[H]
\begin{center}
\resizebox{\linewidth}{!}{\begin{tabulary}{1.12\textwidth}{l|CC|CC|CC|CC}
\hline\hline\rule{0pt}{3ex}
Model  & \multicolumn{2}{c}{$\sigma^{\rm 4FS}_{\rm NLO}(pp\rightarrow tbH^{\pm})$} & \multicolumn{2}{c}{$\sigma^{\rm 5FS}_{\rm NLO}(pp\rightarrow tH^{\pm})$} & \multicolumn{2}{c}{$\sigma^{\rm matched}_{\rm NLO}$} & $\mu_F=\mu_R$ & $\bar{m}_b$\\
  & 14 TeV & 27 TeV & 14 TeV & 27 TeV  & 14 TeV & 27 TeV & & \\ 
\hline\rule{0pt}{3ex} 
\!\!(a) & 49.0$^{+12.6\%}_{-13.1\%}$ & 272.8$^{+9.2\%}_{-10.3\%}$ & 71.8$^{+6.6\%}_{-5.7\%}$ & 397.1$^{+7.0\%}_{-6.6\%}$  & 65.9$^{+8.1\%}_{-7.6\%}$ & 365.4$^{+7.6\%}_{-7.5\%}$ & 183.6 & 2.72  \\
(b)  & 34.5$^{+10.6\%}_{-12.1\%}$ & 204.6$^{+8.1\%}_{-9.6\%}$ & 58.3$^{+7.0\%}_{-5.9\%}$ & 336.1$^{+6.9\%}_{-6.5\%}$ & 52.4$^{+7.9\%}_{-7.4\%}$ & 303.5$^{+7.2\%}_{-7.3\%}$ & 197.9 & 2.70  \\
(c)  & 29.1$^{+11.1\%}_{-12.3\%}$ & 175.9$^{+8.2\%}_{-9.7\%}$ & 48.8$^{+6.7\%}_{-5.7\%}$ & 285.9$^{+6.4\%}_{-6.0\%}$ & 43.9$^{+7.8\%}_{-7.3\%}$ & 259.0$^{+6.8\%}_{-6.9\%}$ & 205.6 & 2.69 \\
(d) & 24.8$^{+10.9\%}_{-12.3\%}$ & 149.9$^{+7.1\%}_{-9.1\%}$ & 42.6$^{+6.3\%}_{-5.3\%}$ & 264.8$^{+6.8\%}_{-6.2\%}$ & 38.3$^{+7.4\%}_{-6.9\%}$ & 237.2$^{+6.8\%}_{-6.9\%}$ & 215.9 & 2.68 \\
(e) & 18.4$^{+11.2\%}_{-12.4\%}$ & 120.1$^{+8.3\%}_{-9.8\%}$ & 32.3$^{+5.9\%}_{-4.9\%}$ & 206.7$^{+6.4\%}_{-6.0\%}$ & 29.0$^{+7.1\%}_{-6.7\%}$ & 186.3$^{+6.8\%}_{-6.9\%}$ & 229.6 & 2.67 \\
(f)  & 13.6$^{+11.3\%}_{-12.5\%}$ & 93.2$^{+7.8\%}_{-9.5\%}$ & 25.1$^{+6.1\%}_{-5.2\%}$ & 169.6$^{+6.7\%}_{-6.0\%}$ & 22.4$^{+7.3\%}_{-6.9\%}$ & 152.1$^{+7.0\%}_{-6.8\%}$ & 248.2 & 2.65  \\
(g) & 13.1$^{+10.5\%}_{-12\%}$ & 95.8$^{+7.6\%}_{-9.5\%}$ & 26.0$^{+6.2\%}_{-5.6\%}$ & 185.1$^{+6.7\%}_{-6.0\%}$ & 23.1$^{+7.2\%}_{-7.0\%}$ & 165.0$^{+6.8\%}_{-6.8\%}$ & 264.6 & 2.64 \\
(h) & 11.2$^{+10.3\%}_{-12.0\%}$ & 85.1$^{+7.5\%}_{-9.4\%}$ & 22.7$^{+6.1\%}_{-5.8\%}$ & 168.3$^{+6.8\%}_{-5.9\%}$ & 20.2$^{+7.0\%}_{-7.2\%}$ & 149.9$^{+6.9\%}_{-6.7\%}$ & 278.2 & 2.63 \\
(i)  & 7.8$^{+11.7\%}_{-12.6\%}$ & 61.1$^{+8.1\%}_{-9.8\%}$ & 15.8$^{+6.0\%}_{-6.0\%}$ & 121.0$^{+6.9\%}_{-6.0\%}$ & 14.0$^{+7.2\%}_{-7.4\%}$ & 107.9$^{+7.2\%}_{-6.8\%}$ & 292.9 & 2.62 \\
(j) & 5.5$^{+12.6\%}_{-13.0\%}$ & 48.9$^{+9.1\%}_{-10.3\%}$ & 11.6$^{+6.7\%}_{-6.5\%}$ & 99.4$^{+6.4\%}_{-5.5\%}$ & 10.3$^{+7.9\%}_{-7.8\%}$ & 88.7$^{+6.9\%}_{-6.5\%}$ & 329.9 & 2.60 \\
\hline
\end{tabulary}}\end{center}
\caption{The NLO production cross-sections, in fb, of the charged Higgs in association with a top (and bottom) quark in the five (and four) flavour schemes along with the matched values at $\sqrt{s}=14$ TeV and $\sqrt{s}=27$ TeV for benchmarks of Table~\ref{tab1}. The running $b$-quark mass, in GeV, is also shown evaluated at the factorization and normalization scales, $\mu_F=\mu_R$ (in GeV).}
\label{tab3}
\end{table}

We give in Table~\ref{tab4} the branching ratios of the dominant charged Higgs decays. The competing channels are $tb$ (first column) and the electroweakinos (last column). In the MSSM, the coupling of the charged Higgs to up-type fermions goes as $\cot\beta$, whereas the couplings to down-type fermions (down quarks and leptons) goes as $\tan\beta$. Typically, the branching ratios of the charged Higgs to $tb$ and $\tau\nu$ must increase with $\tan\beta$ which can be observed in most of the benchmark points. However, there are exceptions since the chargino-neutralino channel is kinematically open. Thus one can see the branching ratios into the electroweakinos vary from about 2\% (point h) to $\sim$ 65\% (point j). For point (h), the only open channel for the electroweakinos is $\chi_1^{\pm}\chi^0_1$ and since the $\mu$ parameter is $\sim 1$ TeV, the Higgsino content of the LSP is small and so is the coupling to the charged Higgs, thus, the observed small branching ratio. However, for point (j), all channels are open, $\chi_1^{\pm}\chi^0_1$, $\chi_1^{\pm}\chi^0_2$, $\chi_1^{\pm}\chi^0_3$, $\chi_1^{\pm}\chi^0_4$, $\chi_2^{\pm}\chi^0_1$, $\chi_2^{\pm}\chi^0_2$, $\chi_2^{\pm}\chi^0_3$ and $\chi_2^{\pm}\chi^0_4$. This point has a small $\mu$ parameter and hence the Higgsino content of the LSP is larger. Indeed, we observe a 6 and 50 fold increase in branching ratios to $\chi_1^{\pm}\chi^0_1$ and $\chi_1^{\pm}\chi^0_2$, respectively. The largest branching ratio, however, comes from $\chi_1^{\pm}\chi^0_4$ and $\chi_2^{\pm}\chi^0_2$ which appear to have a considerable amount of Higgsino content.

\begin{table}[H]
\begin{center}
\begin{tabulary}{0.88\textwidth}{l|CCC}
\hline\hline\rule{0pt}{3ex}
 Model & $H^{+}\rightarrow t\bar{b}$ & $H^{-}\rightarrow \tau\bar \nu_{\tau}$ & $H^{\pm}\rightarrow \sum_{i=1}^2\sum_{j=1}^3 \chi^{\pm}_i\chi^0_j$ \\
\hline\rule{0pt}{3ex} 
\!\!(a) & 0.634 & 0.063 & 0.292 \\
(b) & 0.530 & 0.071 & 0.394 \\
(c) & 0.778 & 0.101 & 0.114 \\
(d) & 0.768 & 0.107 & 0.119 \\
(e) & 0.835 & 0.114 & 0.046 \\
(f) & 0.848 & 0.120 & 0.027 \\
(g) & 0.547 & 0.086 & 0.365 \\
(h) & 0.860 & 0.135 & 0.002 \\
(i) & 0.859 & 0.132 & 0.006 \\
(j) & 0.303 & 0.049 & 0.647 \\
\hline
\end{tabulary}\end{center}
\caption{The branching ratios of the dominant decay channels of the MSSM charged Higgs boson for the benchmarks of Table~\ref{tab1}.}
\label{tab4}
\end{table}

\section{Discovery potential in the $H^{\pm}\rightarrow\tau\nu$ channel \label{sec6}}
We study the discovery prospects of the MSSM charged Higgs boson in its decay to a hadronic tau and missing energy. Theoretical studies of the possibility of observation of a charged Higgs at various colliders are numerous~\cite{Arhrib:2018ewj,Akeroyd:2018axd,Logan:2018wtm,Ahmed:2018jmp,Bahl:2018zmf} including analysis in the $tb$ channel~\cite{Guchait:2018nkp} which showed that a mass range of about 300 to 600 GeV for $\tan\beta=30$ may be observable  with about  $1000\ifb$ of integrated luminosity. The $\tau\nu$ channel has also been studied~\cite{Basso:2015dka,Hashemi:2012xk}. 
Before we discuss the result of our analysis we
describe the different codes utilized  in simulation of the signals and backgrounds. The simulation of the charged Higgs associated production, $t[b]H^{\pm}$, is done at fixed order in NLO using \code{MadGraph} interfaced with LHAPDF~\cite{Buckley:2014ana} and \code{PYTHIA8}~\cite{Sjostrand:2014zea} which handles the showering and hadronization of the samples. The PDF sets used are \code{NNPDF23\_nlo\_FFN\_NF4} for 4FS and \code{NNPDF23\_nlo\_FFN\_NF5} for 5FS at $\alpha_s=0.118$. The charged Higgs branching ratios are calculated by \code{HDECAY}~\cite{Djouadi:1997yw,Djouadi:2018xqq}.

 In the analysis presented here 
 we look at the hadronic tau decay of the charged Higgs accompanied by missing transverse energy (neutrino). This channel has the smallest branching ratio but it is of interest since jets can be tau-tagged and the tau has leptonic and hadronic decay signatures. In the hadronic final states, the tau decay is characterized by its one-prong and three-prong decays which can be utilized to suppress possible SM backgrounds. Hence for such a final state (hadronic tau with missing transverse energy), the SM backgrounds are mainly $t\bar{t}$, $t$+jets, $W/Z/\gamma^*$+jets, diboson production and QCD multijet events which can fake the hadronic tau decays. The backgrounds are simulated at LO using \code{MADGRAPH 2.6.0} with the NNPDF30LO PDF set. The  cross-sections are then normalized to their NLO values. The resulting hard processes are then passed on to \code{PYTHIA8} for showering and hadronization. To avoid double counting of jets, a five-flavor MLM matching~\cite{Mangano:2006rw} is performed on the samples. Jets are clustered with \code{FASTJET}~\cite{Cacciari:2011ma} using the anti-$k_t$ algorithm~\cite{Cacciari:2008gp} with jet radius 0.4. Detector simulation and event reconstruction is performed by \code{DELPHES-3.4.2}~\cite{deFavereau:2013fsa} using the beta card for HL-LHC and HE-LHC studies. The output events are stored in ROOT files and the signal region analysis and processing of those files are done with \code{ROOT 6}~\cite{Antcheva:2011zz}.   

\subsection{Selection criteria}

The selection criteria is based on the flavour scheme under consideration. In the 4FS, the production of a charged Higgs is in association with a bottom and top quarks while the 5FS does not involve a b-quark in the initial state. Hence for the 4FS one has an extra b-tagged jet. We consider the hadronic decay of the top quark ($t\rightarrow b W\rightarrow 3~\text{jets}$, with at least one of the jets being b-tagged) and the hadronic tau ($\tau_h$) decay of the charged Higgs. So we can summarize the selection criteria in the two flavour schemes as
\begin{align}
4\text{FS} &: \text{Lepton veto}, \geq 5j ~(2b, ~1\tau_h), \\
5\text{FS} &: \text{Lepton veto}, \geq 4j ~(1b, ~1\tau_h), 
\end{align} 
where the lepton veto involves rejecting events with electrons and/or muons. The minimum $p_T$ of the leading jet is 20 GeV and that of the tau-tagged jet is 25 GeV.  
We will  discuss  two types of analyses in this work, a cut-based analysis which uses the traditional linear cuts on select kinematic variables and a boosted decision tree analysis, and we will give a relative comparison of these.

\subsection{Cut-based analysis}
We begin the analysis by using a series of linear cuts on select kinematic variables used to discriminate the signal from the background. We carry out the analysis on two benchmark points using 4FS at 14 and 27 TeV. The following set of variables are used for discriminating the signal from the background:

\begin{enumerate}
\item $E^{\rm miss}_T$ and $E^{\rm miss}_T/\sqrt{H_T}$, where the former is the missing transverse energy and the latter is a powerful variable in discriminating against QCD multijet events with $H_T$ being the sum of all $p_T$'s of visible final state particles in an event. 
\item $m_{T2}^{\rm jets}$, the stransverse mass~\cite{Lester:1999tx, Barr:2003rg, Lester:2014yga} of the leading b-tagged and tau-tagged jets defined as
\begin{equation}
    m_{\rm T2}=\min\left[\max\left(m_{\rm T}(\mathbf{p}_{\rm T}^{b},\mathbf{q}_{\rm T}),
    m_{\rm T}(\mathbf{p}_{\rm T}^{\tau},\,\mathbf{p}_{\rm T}^{\text{miss}}-
    \mathbf{q}_{\rm T})\right)\right],
    \label{mt2}
\end{equation}
where $\mathbf{q}_{\rm T}$ is an arbitrary vector chosen to find the appropriate minimum and the transverse mass $m_T$ is given by 
\begin{equation}
    m_{\rm T}(\mathbf{p}_{\rm T1},\mathbf{p}_{\rm T2})=
    \sqrt{2(p_{\rm T1}\,p_{\rm T2}-\mathbf{p}_{\rm T1}\cdot\mathbf{p}_{\rm T2})}.
\end{equation}  
\item $m_T^{\tau}$ and $p_T^{\tau}$, the transverse mass and leading transverse momentum of the hadronic tau. Since $H^{\pm}$ decays to $\tau\nu$, the transverse mass, $m_T^{\tau}$, has a kinematical endpoint at the charged Higgs mass. This proves to be an important discriminant in this analysis.
\item $E^{\rm miss}_T/m_{\rm eff}$, where $m_{\rm eff}$, is the effective mass defined as 
\begin{equation}
m_{\rm eff} = H_T+E^{\rm miss}_T+p^{\tau}_T.
\end{equation}
\end{enumerate}
As a precursor to the analysis based on BDT  in section~\ref{BDT} we first discuss for some sample points  an analysis based on
linear cuts. As an illustration, we consider at 14 TeV the analysis of
benchmark point (c). We present in Fig.~\ref{fig1} distributions in four kinematic variables at an integrated luminosity of $3000\ifb$ with the signal increased 100 folds for clarity.  

\begin{figure}[H]
 \centering
 \includegraphics[width=0.49\textwidth]{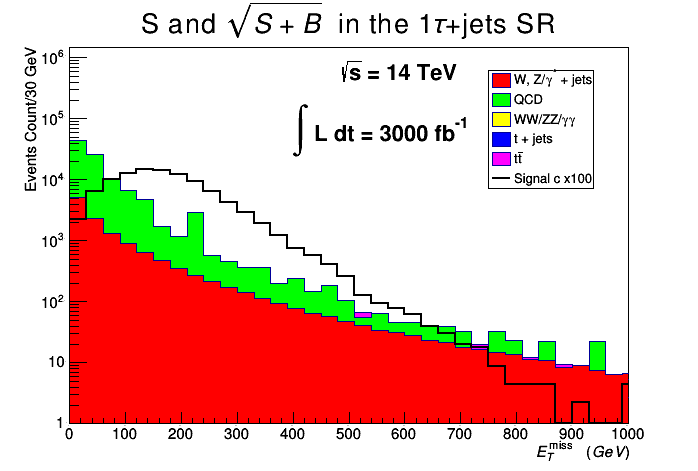} 
 \includegraphics[width=0.49\textwidth]{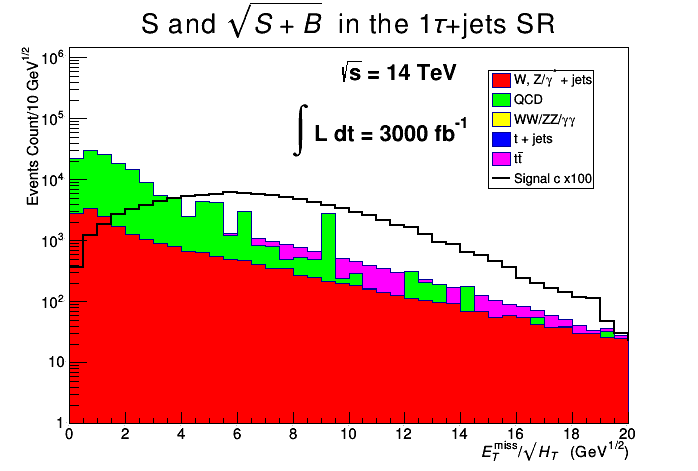}
 \includegraphics[width=0.49\textwidth]{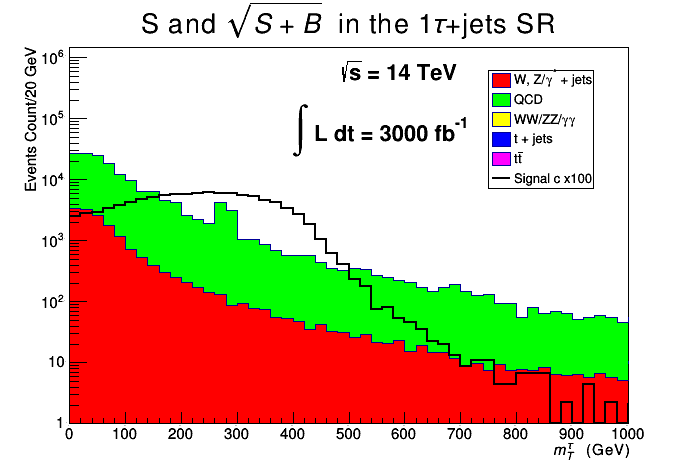} 
 \includegraphics[width=0.49\textwidth]{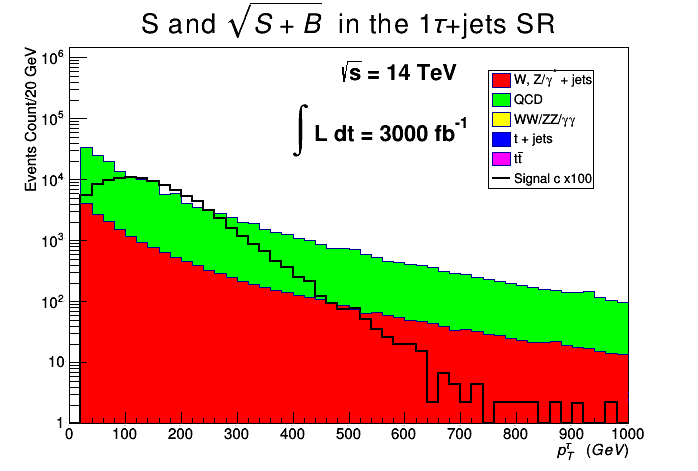}
      
   \caption{{Distributions of signal, S (black histogram), and signal + background, $\sqrt{S+B}$ (color-filled histograms), for four kinematic variables $E^{\rm miss}_T$, $E^{\rm miss}_T/\sqrt{H_T}$, $m^{\tau}_T$ and $p^{\tau}_T$ at 14 TeV for an integrated luminosity of 3000$\ifb$ in the $1\tau+\text{jets}$ signal region (SR).}}
	\label{fig1}
\end{figure}

Table~\ref{tab5} shows a cut-flow for signal and background at 14 TeV. After applying the cuts, $t\bar{t}$ and QCD remain to be significant and a calculation for the required integrated luminosity for a $5\sigma$ discovery gives $\mathcal{O}(10^6)\ifb$ which is beyond the capabilities of the HL-LHC. 

\begin{table}[H]
\begin{center}
\resizebox{\linewidth}{!}{\begin{tabulary}{0.85\textwidth}{l|cccccc}
\hline\hline\rule{0pt}{3ex}
Cuts & Signal & $t\bar{t}$ & $t+$jets & $W/Z/\gamma^*+$ jets&  $WW/ZZ/\gamma\gamma$ & QCD \\
      \hline
Lepton veto, $\geq 5j~(2b~1\tau_h)$ & 0.337 & 20401 & 1309 & 11714 & 277 & $9.58\times 10^5$  \\
$E_{T}^{\text{miss}} > 100$ GeV & 0.262 & 3491 & 122 & 507 & 39 & 18624 \\
$m_T^{\tau} > 200$ GeV & 0.185 & 143 & 3.4 & 23 & 1.6 & 10123 \\
$E^{\rm miss}_T/m_{\rm eff}>0.1$ & 0.178 & 132 & 3.1 & 20 & 1.4 & 9314 \\
$p_T^{\tau}>100$ GeV & 0.143 & 90 & 2.4 & 15 & 0.97 & 9045 \\
$m_{T2}^{\rm jets} > 130$ GeV & 0.142 & 88 & 2.3 & 14.8 & 0.96 & 9000 \\
$E^{\rm miss}_T/\sqrt{H_T}>6$ GeV$^{1/2}$ & 0.106 & 53 & 1.3 & 10 & 0.68 & 2910 \\
\hline
\end{tabulary}}\end{center}
\caption{Cut-flow for signal (point (c) in 4FS) and SM background at $\sqrt{s}=14$ TeV. Samples are normalized to their respective cross-sections (in fb).}
\label{tab5}
\end{table}


A similar analysis is carried out for benchmark point (a) at 27 TeV. We show in Fig.~\ref{fig2} a display of signal and background distributions for four kinematic variables related to benchmark point (a) of Table~\ref{tab1}. The distributions are shown for a 27 TeV center-of-mass energy and an integrated luminosity of 15$\iab$. The signal is increased ten folds to show the best possible cut value for a particular variable. 

\begin{figure}[H]
 \centering
 \includegraphics[width=0.49\textwidth]{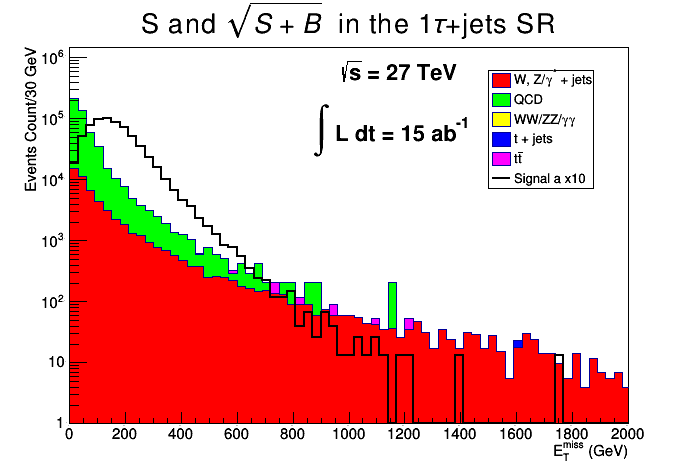} 
 \includegraphics[width=0.49\textwidth]{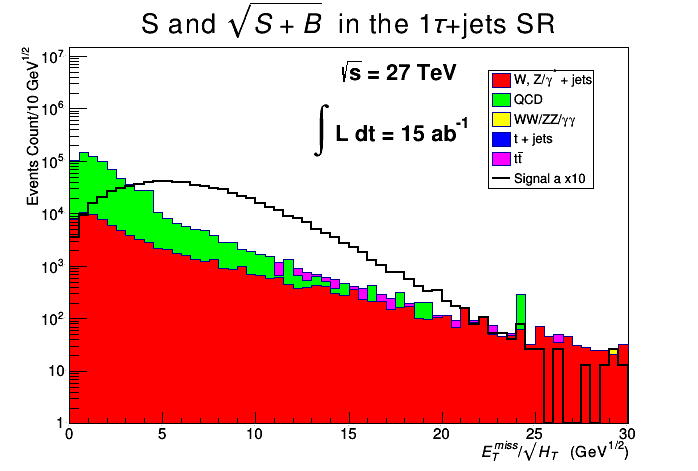}
 \includegraphics[width=0.49\textwidth]{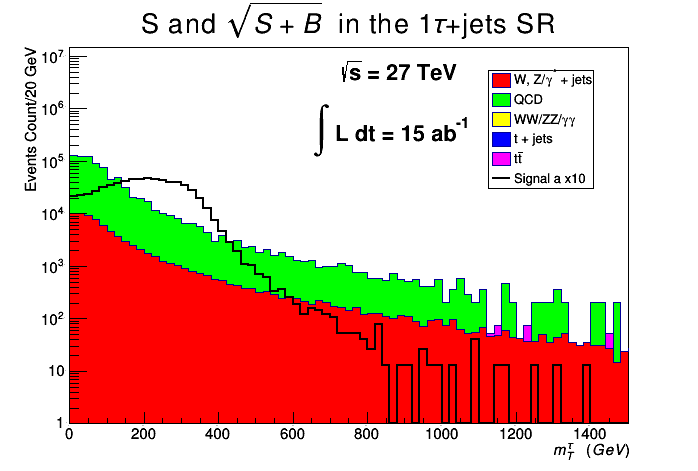} 
 \includegraphics[width=0.49\textwidth]{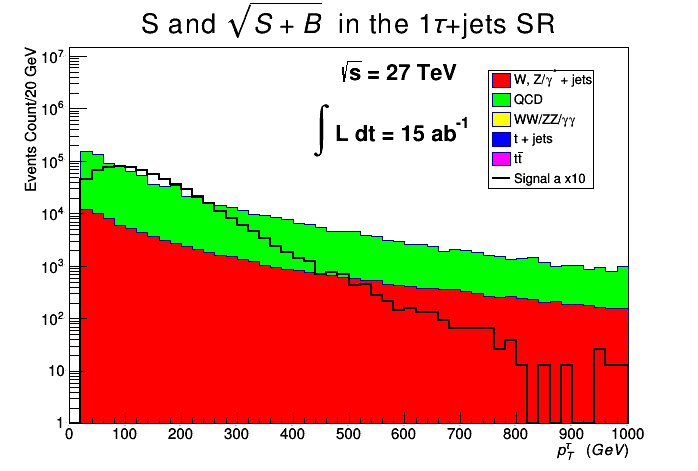}
      
   \caption{{Distributions of signal, S (black histogram), and signal + background, $\sqrt{S+B}$ (color-filled histograms), for four kinematic variables $E^{\rm miss}_T$, $E^{\rm miss}_T/\sqrt{H_T}$, $m^{\tau}_T$ and $p^{\tau}_T$ at 27 TeV for an integrated luminosity of 15$\iab$ in the $1\tau+\text{jets}$ signal region (SR).}}
	\label{fig2}
\end{figure}

Table~\ref{tab6} shows a cut-flow of the signal and background starting with the 4FS selection criteria applied to point (a). One can see that the variables $m_T^{\tau}$, $E^{\rm miss}_T$ and $E^{\rm miss}_T/\sqrt{H_T}$ are effective in reducing the background but not enough to extract the signal. A calculation of the required integrated luminosity {for $\frac{S}{\sqrt{S+B}}$ at the $5\sigma$ level gives} $\sim 60 \iab$ which is way beyond the reach of the proposed 27 TeV hadron collider.    

\begin{table}[H]
\begin{center}
\resizebox{\linewidth}{!}{\begin{tabulary}{0.85\textwidth}{l|cccccc}
\hline\hline\rule{0pt}{3ex}
Cuts & Signal & $t\bar{t}$ & $t+$jets & $W/Z/\gamma^*+$ jets&  $WW/ZZ/\gamma\gamma$ & QCD \\
      \hline
Lepton veto, $\geq 5j~(2b~1\tau_h)$ & 4.40 & 77383 & 4764 & 30597 & 822 & $4.63\times 10^6$  \\
$E_{T}^{\text{miss}} > 100$ GeV & 3.22 & 12992 & 529 & 2393 & 153 & 54845 \\
$m_T^{\tau} > 130$ GeV & 2.63 & 2555 & 100 & 737 & 36 & 27648 \\
$E^{\rm miss}_T/m_{\rm eff}>0.1$ & 2.52 & 2115 & 79 & 528 & 26 & 17730 \\
$p_T^{\tau}>100$ GeV & 1.56 & 778 & 33 & 231 & 12 & 7592 \\
$m_{T2}^{\rm jets} > 130$ GeV & 1.54 & 755 & 32 & 229 & 12 & 7348 \\
$E^{\rm miss}_T/\sqrt{H_T}>6$ GeV$^{1/2}$& 1.03 & 350 & 14 & 104 & 7 & 2081 \\
\hline
\end{tabulary}}\end{center}
\caption{Cut-flow for signal (point (a) in 4FS) and SM background at $\sqrt{s}=27$ TeV. Samples are normalized to their respective cross-sections (in fb).}
\label{tab6}
\end{table} 

Here we note that previous analyses on $H^{\pm}\rightarrow\tau\nu$ channel used the $\tau$ polarization (one and three-prong decays)~\cite{Raychaudhuri:1995cc,Guchait:2008ar,Guchait:2006jp} which showed that a signal may be extracted for up to $100\ifb$ of integrated luminosity for a mass range of 200-800 GeV in the moderate and high $\tan\beta$ ranges. Most of those points have already been excluded by ATLAS and CMS. Nevertheless, the technique showed success in suppressing $t\bar{t}$ and QCD backgrounds (for further discussion of this topic and of other techniques see~\cite{Bullock:1991fd,Roy:1991sf,Boos:2005ca,Tanaka:2010se}). 
Despite the successes of many of those techniques one can still face trouble especially in low mass regime where the final states of the charged Higgs decay look very much like the SM background especially with the presence of hadronic tau fakes from QCD. In order to refine the search for the charged Higgs we resort to machine learning techniques which are taking center stage in high energy physics in data analysis.    

\subsection{Analysis using Boosted Decision Trees \label{BDT}}

Boosted decision trees  have been around and used in high energy physics for some time now. Analyses based on BDTs have appeared in searches by ATLAS and CMS collaborations, and helped in the discovery of the SM Higgs boson and more recently in the observation of the decay $h\rightarrow b\bar{b}$~\cite{Aaboud:2018zhk} (In fact ATLAS used BDTs in this analysis while CMS used another machine learning technique known as deep neural network~\cite{CMS:2018abb}). BDTs prove to be very helpful in separating signal and background especially for the cases where the signal is small and the background is overwhelming such that simple linear cuts fail to be of much help in such an environment.
Based on multivariate analysis technique, BDTs use a set of kinematic variables to make a decision on whether an event is to be classified as a signal or a background. At the end of the training process, a single variable is created, known as the BDT response or score, which is used as a discriminator.

BDTs consist of many decision trees that constitute a series of ``weak learners"~\cite{Barber:2006yq} and based on multivariate analysis technique which make them powerful tools for classification problems. A tree consists of nodes and leaves which all ramify from the main node called a root node.  {The node refers to a criterion set on a variable which can be a ``pass" or ``fail"}.  The training of the trees starts with the algorithm selecting a variable which best separates the signal from the background. A cut value of this variable is chosen and applied to the events which are split into left or right nodes depending whether they are classified as signal or background. A new variable is chosen with the best cut to further split the data into signal or background. The splitting into nodes ends when the maximum depth of the tree is reached or some stopping citeria is given. The tree ends with leaves where events classified as signal are assigned the value +1 and a value of $-1$ if classified as background. Misclassified events, i.e. signal events that end up in background nodes and vice-versa, are given larger weights and the whole process starts again with a new root node. The reason for providing extra weights to those events is that in the next iteration, more attention will be paid to those events and separation efficiency becomes better. More trees are created until the grown "forest" have the specified number of trees and the training process ends. The next process is the testing process to check how well the BDTs have learned about the signal and background features. The testing is done on a separate Monte Carlo sample so that the training and testing processes are statistically independent. The end result of the testing phase is the BDT score variable which can be used as a discriminating variable. An important issue to be aware of is overtraining. The performance of the BDT in the testing phase should not outdo that of the training phase. This can happen in some cases where BDT classifies events according to some specific features found in the training sample. Overtraining can be avoided by controlling the number of trees to be trained and their maximum depth. Usually a choice of a maximum depth of more than 4 on a sample with not enough statistics will result in overtraining.    

        The type of BDT we use in this analysis is known as gradient boosted decision tree,\\
 \code{GradientBoost}. The main differences between the various kinds of BDTs lie in the loss functions used. \code{GradientBoost} uses a binomial log-likelihood loss function which is ideal for weak classifiers, i.e. trees with a depth of 2 to 4. The effectiveness of \code{GradientBoost} can be enhanced by reducing the learning rate using the \code{Shrinkage} parameter which was set to 0.2 in this analysis. The number of trees in the forest ranged between 600 to 1500 and the maximum depth between 3 and 4 depending on whether enough statistics is present in the samples or not. With each choice of the number of trees and maximum depth we made sure no overtraining was present. 
A large set of variables have been tried and the ones which produced the best results were kept and used for all the signal points (a)-(j). The kinematic variables used in the previous section along with the ones listed here enter into the training of the BDTs: 

\begin{enumerate}

\item The minimum transverse mass, $m_T^{\rm min}(j_{1-2},E^{\rm miss}_T)$, of the two leading untagged jets.  
This variable is effective in reducing $t\bar{t}$, $W+$ jets and QCD multijet backgrounds. 
 
\item $\Delta\phi(p^{\tau}_T, E^{\rm miss}_T)$, the opening angle between the leading hadronic tau and missing transverse momentum. This variable tends to be larger for the signal, i.e. $\gtrsim 1.5$ rad. 
\item $\ln p_T$, the logarithm of the leading jet $p_T$ if present and zero if no jet exists. 
\item The number of tracks associated with the hadronic tau decay, $N_{\rm tracks}^{\tau}$. It is a very effective variable which enables the BDT to differentiate between tau decays {according to their charge multiplicities since tau decays can proceed as  one prong  or three-prong decays}. 
\item $\sum_{\rm tracks} p_T$, the sum of the track $p_T$'s.

\end{enumerate}

The training and testing of the samples is carried out using ROOT's own TMVA (Toolkit for Multivariate Analysis) framework~\cite{Speckmayer:2010zz}. In the training of the BDTs, the algorithm ranks the variables in decreasing order of importance. The variable which is ranked at the top is the one the BDT has used the most during the training in order to separate the signal from the background. The ranking of the variables differs from one point to another especially between the ones with very different charged Higgs mass. For this reason, we can split the benchmark points we have into two sets, one with low charged Higgs mass, i.e. $m_{H^{\pm}}<500$ GeV (points (a)-(e)), and another with high charged Higgs mass, i.e. $m_{H^{\pm}}>500$ GeV (points (f)-(j)). We present in Table~\ref{tab7} the ranking of the variables for the two charged Higgs mass ranges. 

\begin{table}[H]
	\centering
	\begin{tabulary}{\linewidth}{l|c|c}
    \hline\hline
	Rank & Low mass range & High mass range \\
	\hline
  (1) & $E^{\rm miss}_T$ & $m_{\rm eff}$ \\
  (2) & $m_{T2}^{\rm jets}$ & $m_{T2}^{\rm jets}$ \\ 
  (3) & $E^{\rm miss}_T/\sqrt{H_T}$ & $H_T$ \\
  (4) & $E^{\rm miss}_T/m_{\rm eff}$ & $E^{\rm miss}_T$ \\ 
  (5) & $m_T^{\tau}$ & $\ln p_T$ \\ 
  (6) & $m_{\rm eff}$ & $E^{\rm miss}_T/\sqrt{H_T}$ \\ 
  (7) & $N_{\rm tracks}^{\tau}$ & $m_T^{\rm min}(j_{1-2},E^{\rm miss}_T)$ \\ 
  (8) & $m_T^{\rm min}(j_{1-2},E^{\rm miss}_T)$ & $E^{\rm miss}_T/m_{\rm eff}$ \\ 
  (9) & $p_T^{\tau}$ & $m_T^{\tau}$ \\ 
  (10) & $\ln p_T$ & $N_{\rm tracks}^{\tau}$ \\ 
  (11) & $H_T$ & $p_T^{\tau}$ \\
  (12) & $\sum_{\rm tracks} p_T$ & $\sum_{\rm tracks} p_T$ \\
  (13) & $\Delta\phi(p^{\tau}_T, E^{\rm miss}_T)$ & $\Delta\phi(p^{\tau}_T, E^{\rm miss}_T)$ \\
	\hline
	\end{tabulary}
	\caption{The ranking of variables entering in the training of the BDTs in decreasing order of importance for the two charged Higgs mass ranges. }
\label{tab7}
\end{table}
   
After the training and testing phase, the variable ``BDT score" is created. Cuts on this variable will allow us to eliminate most of the background events. However, this is not enough. In addition to the selection criteria discussed in the previous section, additional cuts on some variables are necessary to extract the signal. Those cuts vary from one point to another and are summarized in Table~\ref{tab8}.   

\begin{table}[H]
	\centering
	\begin{tabulary}{\linewidth}{l|ccc}
    \hline\hline
	Model & $E^{\rm miss}_T/\sqrt{H_T}$ (GeV$^{1/2}$)& $m_{\rm eff}$ (GeV) & $m_T^{\rm min}(j_{1-2},E^{\rm miss}_T)$ (GeV) \\
	\hline
  (a) & 3 &  &  \\
  (b) & 7 & 500 & 100 \\
  (c) & 11 & 500 & 95 \\
  (d) & 11 & 500 & 100 \\
  (e) & 7 & 500 & 100 \\
  (f) & 9 & 500 & 100 \\
  (g) & 9 & 600 &  \\
  (h) & 9 & 1000 & 100 \\
  (i) & 11 & 1000 & 100 \\
  (j) & 11 &  &  \\
	\hline
	\end{tabulary}
	\caption{Additional selection criteria for the ten benchmark points. These consist of lower limits on $E^{\rm miss}_T/\sqrt{H_T}$ shown in column two, on $m_{\rm eff}$ shown in column three and on $m_T^{\rm min}(j_{1-2},E^{\rm miss}_T)$ shown in column four. Empty entries in the table indicate that the cuts in these cases were not effective in reducing the background when coupled with cuts on the BDT score.}
\label{tab8}
\end{table}

We present in Figs.~\ref{fig3}-\ref{fig4} the distributions of signal and background in the BDT score variable for points (a) and (d) after all selection criteria have been applied. The signal is more concentrated close to a BDT score of 1 as one would expect. A comparison between the 14 and 27 TeV cases for the same integrated luminosities of 300 and 1000 $\ifb$ shows the signal to be above the background for a BDT score $>0.95$ only in the 27 TeV case.  

\begin{figure}[H]
 \centering
 \includegraphics[width=0.49\textwidth]{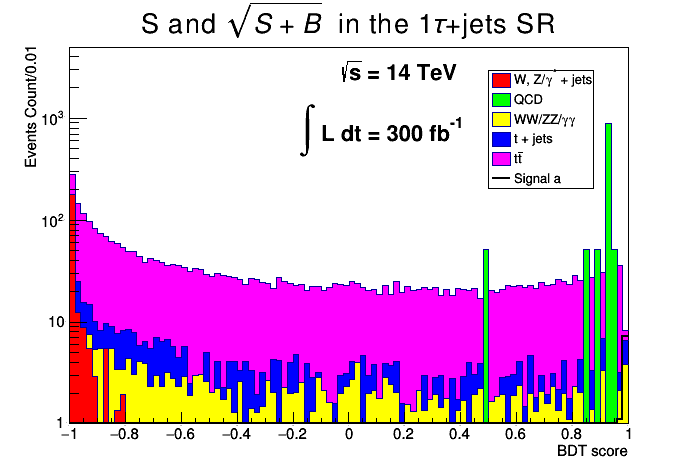} 
  \includegraphics[width=0.49\textwidth]{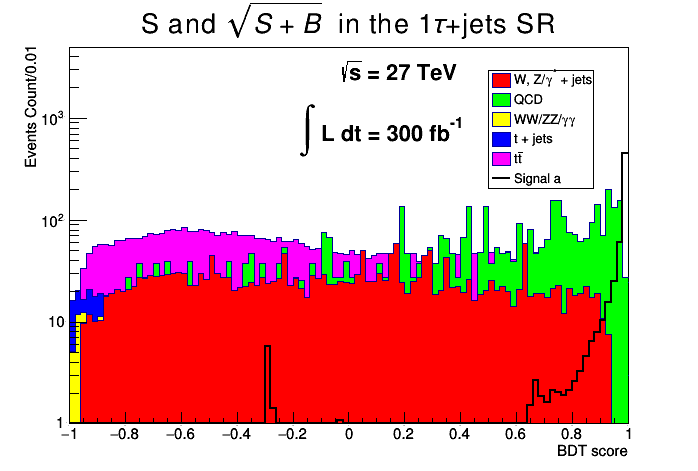}
  \includegraphics[width=0.49\textwidth]{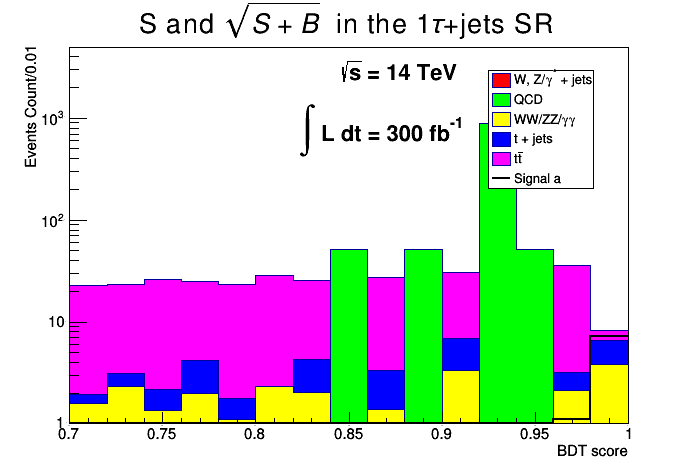} 
  \includegraphics[width=0.49\textwidth]{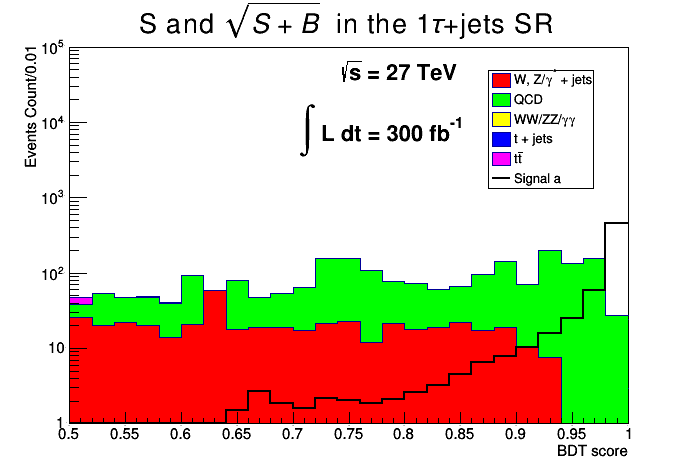}
   \caption{Top row: The BDT score for benchmark points (a) at 14 TeV and 27 TeV in the hadronic tau channel. Bottom row: same plots as the top ones but zoomed in to show signal more clearly.}
	\label{fig3}
\end{figure}

\begin{figure}[H]
 \centering
   \includegraphics[width=0.49\textwidth]{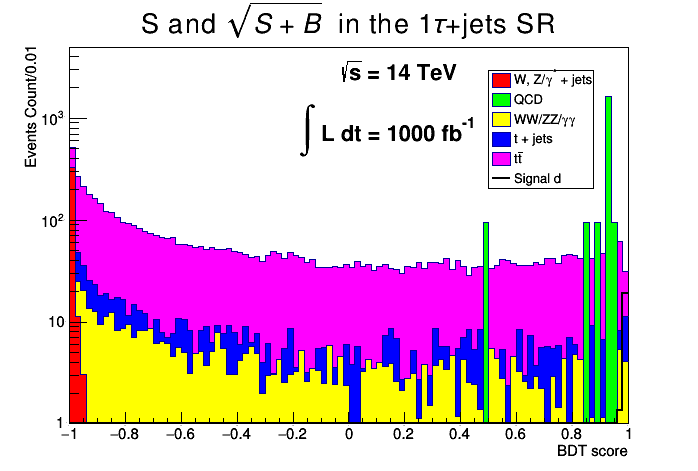} 
      \includegraphics[width=0.49\textwidth]{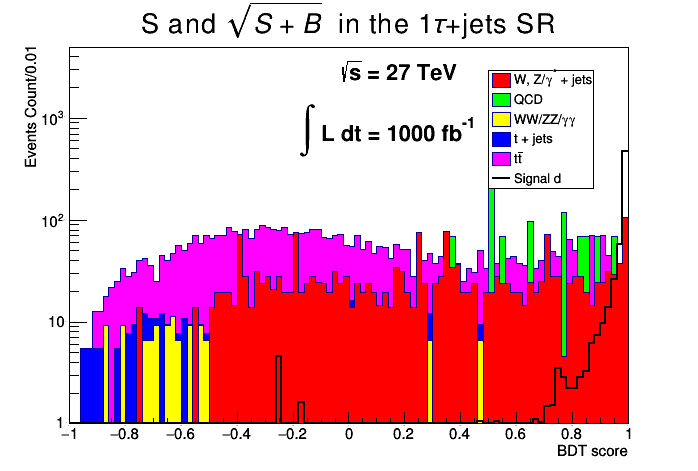}
   \caption{The BDT score for benchmark points (d) at 14 TeV and 27 TeV in the hadronic tau channel.}
	\label{fig4}
\end{figure}
{Note that the spikes in the signal appearing at around a BDT score of -0.2 and -0.3 can be atrributed to statistical fluctuations resulting from the training and testing phase. Those are events that are misclassified as background and given a BDT score $< 0$.} 

The cuts on the BDT score ranges between 0.9 up to 0.98 for the different benchmark points. Fig.~\ref{fig5} shows how the estimated integrated luminosity varies as a function of the BDT score cuts for the different benchmark points at 27 TeV. Starting with a very high integrating luminosities for cuts between -1 and 0, we start seeing a drop for cuts $>0.6$ until a major dip is observed for values $>0.9$. A zoomed in plot (on the right) shows major activity happening between 0.95 and 1 where the dip occurs. The integrated luminosity shoots back up when the cut becomes too strong that no more signal events survive.  

\begin{figure}[H]
 \centering
   \includegraphics[width=0.49\textwidth]{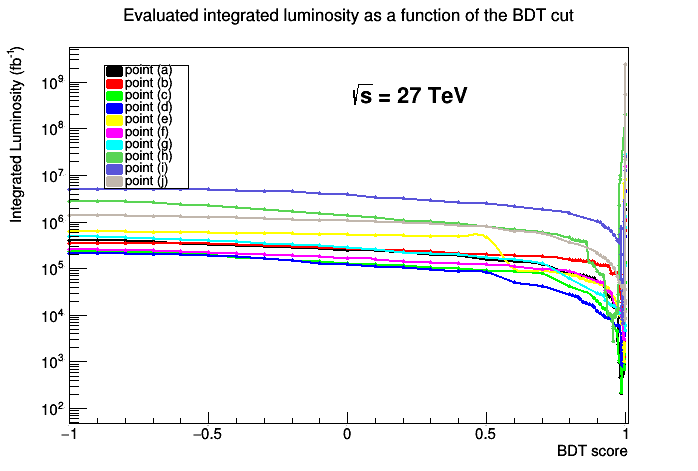} 
      \includegraphics[width=0.49\textwidth]{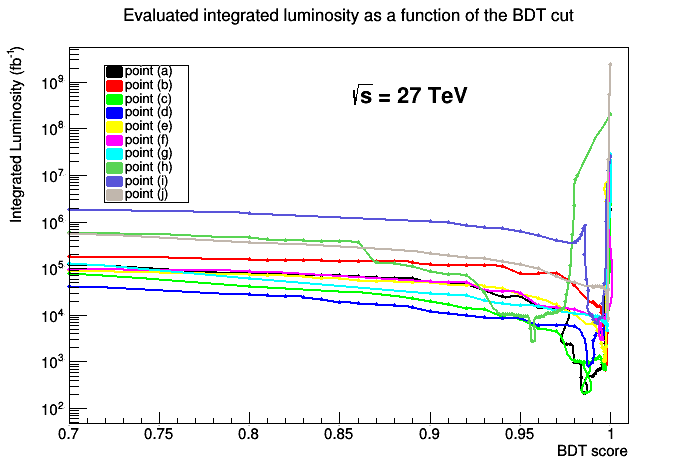}
   \caption{The calculated integrated luminosities as a function of the BDT cut for the ten benchmark points at 27 TeV.}
	\label{fig5}
\end{figure}

We apply the selection criteria along with a BDT score cut $>0.95$ on the SM background and on each of the 4FS and 5FS signal samples to obtain the remaining cross-sections. The signal cross-sections are combined using Eq.~(\ref{matched}) in order to evaluate the required minimum integrated luminosity {for $\frac{S}{\sqrt{S+B}}$ at the $5\sigma$ level} discovery. The results for both the 14 and 27 TeV cases are shown in Table~\ref{tab9}. 

\begin{table}[H]
	\centering
	\begin{tabulary}{\linewidth}{l|c|c}
    \hline\hline
    & \multicolumn{2}{c}{$\mathcal{L}$ for $5\sigma$ discovery in 1$\tau_h$ + jets}  \\
	\hline
	Model & $\mathcal{L}$ at 14 TeV & $\mathcal{L}$ at 27 TeV \\
	\hline
  (c) & 2322 & 219 \\
  (a) & 2387 & 213 \\ 
  (d) & 2908 & 982 \\
  (b) & 3218 & 988 \\ 
  (e) & ... & 1018 \\ 
  (h) & ... & 2750 \\ 
  (f) & ... & 3040 \\ 
  (g) & ... & 4675 \\ 
  (i) & ... & 6636 \\ 
  (j) & ... & 8379 \\ 
	\hline
	\end{tabulary}
	\caption{Comparison between the estimated integrated luminosity ($\mathcal{L}$)
	for a 5$\sigma$ discovery at 14 TeV (middle column)
	 and 27 TeV (right column)
		 for the charged Higgs following the selection cuts and BDT $> 0.95$, where the minimum integrated luminosity needed for a $5\sigma$ discovery is given in fb$^{-1}$. Entries with $\cdots$ mean that the evaluated $\mathcal{L}$ is much greater than $3000\ifb$.}
\label{tab9}
\end{table}
One can see from Table~\ref{tab9} that four of the ten points may be discoverable at the HL-LHC as it nears the end of its run where a maximum integrated luminosity of 3000$\ifb$ will be collected. Given the rate at which the HL-LHC will be collecting data, points (a)-(d) will require $\sim 7$ years of running time. On the other hand, the results from the 27 TeV collider show that all points may be discoverable for integrated luminosities much less than $15\iab$. The HE-LHC will be collecting data at a rate of $\sim 820\ifb$ per year and with that points (a) and (c) may be discoverable with in the first 3 months of operation, points (b), (d) and (e) may take $\sim 1.2$ years, points (h) and (f) $\sim 3.5$ years and the rest of the points $>6$ years.  In the analysis we have not included the effects of CP phases which can be large in supergravity models 
 and can have in general significant effect on phenomena consistent with electric dipole moment constraints (see, e.g., 
 \cite{cp}). It should be of interest, however, to investigate such effects at HL-LHC and HE-LHC in a future work.

\section{Dark matter direct detection\label{sec7}}
Finally we discuss constraints from the direct detection of dark matter experiments in the model. 
For most of the parameter points of Table~\ref{tab1}, $\mu$ is small which renders an LSP with a considerable amount of Higgsino content as shown in Table~\ref{tab10}. A Higgsino LSP has a large spin-independent (SI) proton-neutralino cross-section which puts strong constraints on the model from dark matter direct detection experiments~\cite{Akerib:2016vxi,Cui:2017nnn,Aprile:2018dbl}.
 Thus recent results from XENON1T~\cite{Aprile:2018dbl} show a sensitivity in the SI $p$-$\chi^0_1$ cross-section reaching just below $\mathcal{O}(10^{-46}$) cm$^2$ for an LSP mass less than 100 GeV and rises above that value for masses greater than 100 GeV (see Fig.~\ref{fig6}). Projected sensitivity for XENONnT may reach $\mathcal{O}(10^{-47}$) cm$^2$ in the near future. The neutralino LSP is a mixture of bino, wino and higgsinos such that $\tilde{\chi}_0=\alpha\lambda^0+\beta\lambda^3+\gamma\tilde{H}_1+\delta \tilde{H}_2$, where $\alpha$ is the bino content, $\beta$ is the wino content and $\sqrt{\gamma^2+\delta^2}$ is the higgsino content of the LSP. In Table~\ref{tab10} we give the individual contents of the LSP along with the SI proton-neutralino cross-section, $\mathcal{R}\times \sigma_{SI}$, {where $\mathcal{R}=(\Omega h^2)_{\tilde{\chi}^0_1}/(\Omega h^2)_{\rm PLANCK}$} such that $(\Omega h^2)_{\rm PLANCK}$ is given by Eq.~(\ref{relic}).  

\begin{table}[H]
\begin{center}
\begin{tabulary}{1.3\textwidth}{l|CCCC}
\hline\hline\rule{0pt}{3ex}
Model  & $|\alpha|$ & $|\beta|$ & $\sqrt{\gamma^2+\delta^2}$ & $\mathcal{R}\times\sigma_{SI}$ (cm$^2$)\\
\hline\rule{0pt}{3ex} 
\!\!(a) & 0.065 & 0.810 & 0.580 & $9.70\times 10^{-47}$ \\
(b) & 0.068 & 0.893 & 0.445 & $1.05\times 10^{-46}$ \\
(c) & 0.153 & 0.957 & 0.286 & $6.89\times 10^{-47}$ \\
(d) & 0.136 & 0.959 & 0.247 & $5.62\times 10^{-47}$ \\
(e) & 0.305 & 0.942 & 0.140 & $2.99\times 10^{-47}$ \\
(f) & 0.233 & 0.967 & 0.106 & $1.26\times 10^{-47}$ \\
(g) & 0.190 & 0.917 & 0.352  & $1.18\times 10^{-46}$ \\
(h) & 0.995 & 0.077 & 0.018  & $1.48\times 10^{-46}$ \\
(i) & 0.972 & 0.230 & 0.039  & $3.81\times 10^{-47}$ \\
(j) & 0.175 & 0.927 & 0.331 & $6.55\times 10^{-47}$ \\
\hline
\end{tabulary}\end{center}
\caption{The bino, wino and higgsino content of the neutralino LSP along with the SI proton-neutralino scattering cross-sections for the benchmark points of Table~\ref{tab1}.}
\label{tab10}
\end{table}
In Fig.~\ref{fig6}  the ten benchmark points  of Table~\ref{tab10} are overlaid on the exclusion plot of~\cite{Aprile:2018dbl} and appear
in the red box. Here one finds that all points appear to lie  close to but below the XENON1T upper limit. Thus improved experiment
can either discover dark matter or eliminate some of the parameter points on the plot. We emphasize again that neutralino content of 
dark matter in the model is typically small order a percentage or less, and thus bulk of the dark matter must have a different 
source such as ultralight dark axion mentioned earlier~\cite{Hui:2016ltb,Halverson:2017deq}.
  The fact that the neutralino content of dark matter is small  also appears in other recent models with small $\mu$  discussed in~\cite{Baer:2018avn,Baer:2018hpb,Baer:2018rhs}.

\begin{figure}[H]
 \centering
   \includegraphics[width=0.80\textwidth]{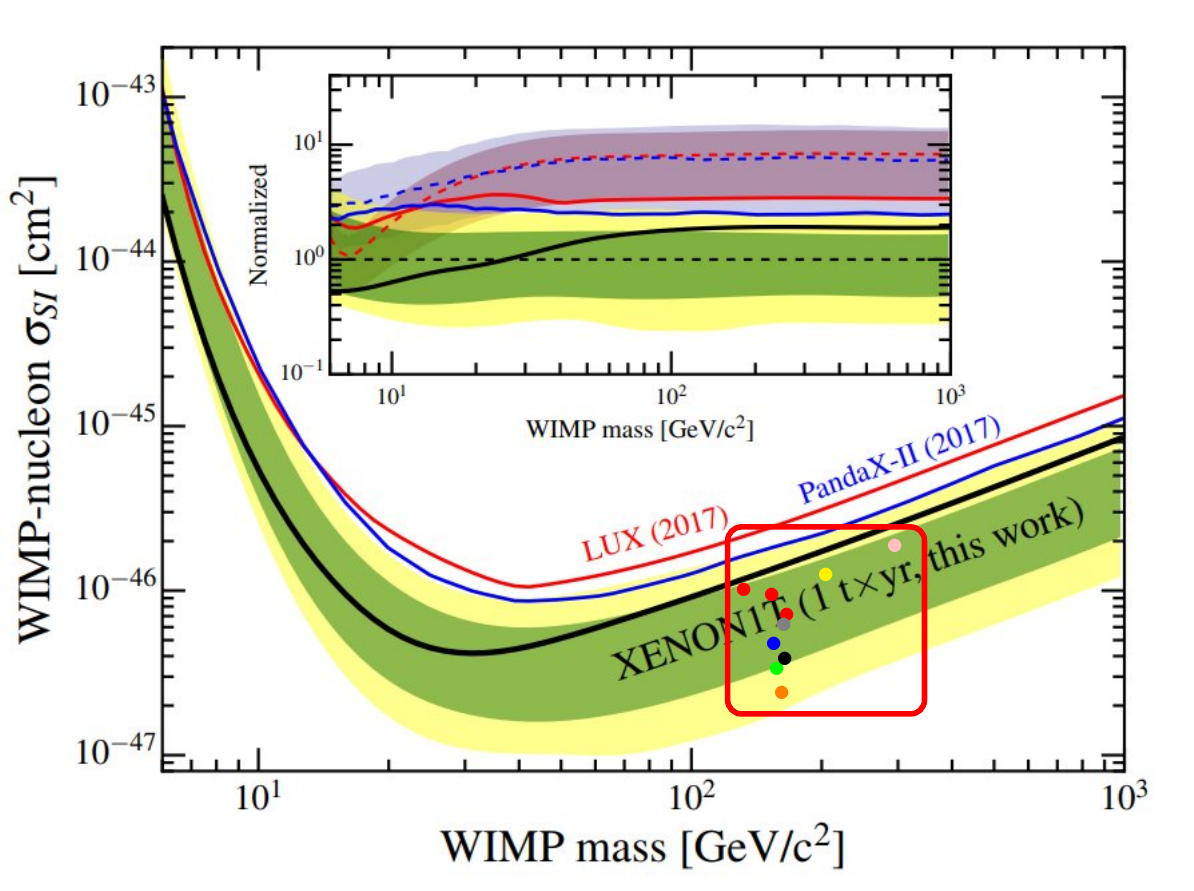} 
   \caption{The SI proton-neutralino cross-section exclusion limits as a function of the LSP mass from XENON1T results
   taken from~\cite{Aprile:2018dbl}.   
    The ten benchmark points are overlaid on the plot showing them lying below but close to the upper limit (black curve).  {The inset shows the limits from LUX 2017, PandaX-II and XENON1T along with the uncertainty bands normalized to the sensitivity median defined in~\cite{Aprile:2018dbl}.}}
	\label{fig6}
\end{figure}


\section{Conclusions\label{sec8}}
In this paper we have given an analysis of the  potential of HL-LHC and HE-LHC for the discovery of the 
charged Higgs boson in the $\tau\nu$ channel  for a mass range of 370-800 GeV for  moderate values of $\tan\beta$  using machine learning technique of boosted decision trees.
 It is shown that the use of machine learning technique allows one to differentiate a signal from the background more efficiently and
thus discover models which would otherwise  not be discoverable using traditional linear cuts.
It is found that using BDTs, charged Higgs with a mass in the range $\sim (370- 470)$ GeV and  $\tan\beta$ in the range 8 to 11 (benchmarks (a)-(d)) may be discoverable at the HL-LHC with an integrated luminosity in the range  $\sim (2300-3000)\ifb$. The same analysis is carried out at 27 TeV for the HE-LHC and it is found that all of the ten benchmark points may be discoverable with an integrated luminosity 
as low  as $\sim 200\ifb$ for point (c) and up to $\sim 8000\ifb$ for point (j). Based on the rate at which data will be collected at the HL-LHC and HE-LHC it is  found that for points (a)-(d) which are discoverable at both machines, one requires a run of $\sim 7$ years at the HL-LHC whereas the run time drops to a few months at the HE-LHC. For the remaining parameter points which are only discoverable at the HE-LHC, a run time ranging from one  year to more than 6 years may be required for the higher mass ranges. These results suggest that a transition from
HL-LHC to HE-LHC when technologically feasible 
would significantly expedite  the 
discovery  of   the charged Higgs in the mass range considered in this work.  {The analysis was done in the context of the SUGRA-MSSM model and thus exclusion limits from ATLAS and CMS pertinent to this class of models were used. Discovery of charged Higgs was studied also in 
{models such as} hMSSM and 2HDM.  Further, other charged Higgs decay channels such as $tb$ and electroweak gauginos are all interesting and require separate analyses. Regarding the $tb$ channel, this has the leading branching ratio for the benchmark points under consideration. However, signatures investigated for this decay mode are often not very successful owing to the difficulty in separating the signal from $t\bar{t}$ background. For this reason, the mode $H^{\pm}\rightarrow\tau\nu$ is favorable because of its cleaner signature.}
Finally we 
 note that the observation of a charged Higgs  boson 
    with a mass much less than $m_0$ would point to the hyperbolic branch where radiative breaking of the electroweak
    symmetry occurs. Further, if the {charged Higgs boson} mass is seen to lie in the few hundred GeV range, then such an 
    observation would lend support to the idea of 
         naturalness defined by small {MSSM Higgs mixing parameter $\mu$.}

\textbf{Acknowledgments:}
The analysis presented here was done using the resources of the high-performance  Cluster353 at the Advanced Scientific Computing Initiative (ASCI) and the Discovery Cluster at Northeastern University. We would like to thank Yacine Haddad at CERN for useful discussions on boosted decision trees. This research was supported in part by the NSF Grant PHY-1620575.

\end{document}